\documentclass[aps,prl,twocolumn,groupedaddress]{revtex4-1}

\usepackage{graphicx,color,graphics}
\usepackage{amsmath}
\usepackage{amssymb}
\usepackage{amsfonts}
\usepackage{hyperref}
\usepackage{url}
\usepackage{bm}
\usepackage{xcolor}

\makeatletter

\newcommand{\ket}[1]{{{|}{#1}\rangle}}
\newcommand{\bra}[1]{{\langle{#1}{|}}}

\begin{document}

\title{High-Fidelity Universal Gate Set for $^9$Be$^+$ Ion Qubits}

\author{J. P. Gaebler}
\author{T. R. Tan}
\email[Electronic address: ]{tingrei.tan@nist.gov}
\author{Y. Lin}
\altaffiliation{Current address: Physics Department and JILA, University of Colorado Boulder, Boulder, Colorado, USA. }
\author{Y. Wan}
\author{R. Bowler}
\altaffiliation{Current address: Physics Department, University of Washington, Seattle, Washington, USA. }
\author{A. C. Keith}
\author{S. Glancy}
\author{K. Coakley}
\author{E. Knill}
\author{D. Leibfried}
\author{D. J. Wineland}
\affiliation{National Institute of Standards and Technology, 325 Broadway, Boulder, CO 80305, USA}

\begin{abstract}
We report high-fidelity laser-beam-induced quantum logic gates on magnetic-field-insensitive qubits comprised of hyperfine states in $^{9}$Be$^+$ ions with a memory coherence time of more than 1 s.  We demonstrate single-qubit gates with error per gate of $3.8(1)\times 10^{-5}$. By creating a Bell state with a deterministic two-qubit gate, we deduce a gate error of $8(4)\times10^{-4}$. We characterize the errors in our implementation and discuss methods to further reduce imperfections towards values that are compatible with fault-tolerant processing at realistic overhead.
\end{abstract}


\maketitle

Quantum computers can solve certain problems that are thought to be intractable on conventional computers.  An important general goal is to realize universal quantum information processing (QIP), which could be used for algorithms having a quantum advantage over processing with conventional bits as well as to simulate other quantum systems of interest \cite{Feynman1982,Deutsch1985,Lloyd1996}.  For large problems, it is generally agreed that individual logic gate errors must be reduced below a certain threshold, often taken to be around $10^{-4}$ \cite{Preskill1998,Knill2010,Ladd2010}, to achieve fault tolerance without excessive overhead in the number of physical qubits required to implement a logical qubit. This level has been achieved in some experiments for all elementary operations including state preparation and readout, with the exception of two-qubit gates, emphasizing the importance of improving multi-qubit gate fidelities. 

Trapped ions are one candidate for scalable QIP.  State initialization, readout, and quantum logic gates have been demonstrated in several systems with small numbers of trapped ions using various atomic species including $^{9}$Be$^+$, $^{25}$Mg$^+$, $^{40}$Ca$^+$, $^{43}$Ca$^+$, $^{88}$Sr$^+$, $^{111}$Cd$^+$, $^{137}$Ba$^+$, and $^{171}$Yb$^+$. The basic elements of scalable QIP have also been demonstrated in multi-zone trap arrays \cite{Home2009,Hanneke2010}. As various ions differ in mass, electronic, and hyperfine structure, they each have technical advantages and disadvantages.  For example, $^{9}$Be$^+$ is the lightest ion currently considered for QIP, and as such, has several potential advantages. The relatively light mass yields deeper traps and higher motional frequencies for given applied potentials, and facilitates fast ion transport \cite{Bowler2012,Walther2012}. Light mass also yields stronger laser-induced effective spin-spin coupling (inversely proportional to the mass), which can yield less spontaneous emission error for a given laser intensity \cite{Ozeri2007}.  However, a disadvantage of $^{9}$Be$^+$ ion qubits compared to some heavier ions such as $^{40}$Ca$^+$ and $^{43}$Ca$^+$ \cite{Benhelm2008,Ballance2015} has been the difficulty of producing and controlling the ultraviolet (313 nm) light required to drive $^{9}$Be$^+$ stimulated-Raman transitions. In the work reported here, we use an ion trap array designed for scalable QIP \cite{Blakestad2011} and take advantage of recent technological developments with lasers and optical fibers that improve beam quality and pointing stability. We also implement active control of laser pulse intensities to reduce errors.  We demonstrate laser-induced single-qubit computational gate errors of $3.8(1) \times 10^{-5}$ and realize a deterministic two-qubit gate to ideally produce the Bell state  $\ket{\Phi_+} = \textstyle{\frac{1}{\sqrt{2}}}(\ket{\uparrow\uparrow}+\ket{\downarrow\downarrow})$. By characterizing the effects of known error sources with numerical simulations and calibration measurements, we deduce an entangling gate infidelity or error of $\epsilon = 8(4)\times10^{-4}$, where $\epsilon =$ 1 - F, and F is the fidelity.  Along with Ref. \cite{Ballance2015}; these appear to be the highest two-qubit gate fidelities reported to date. 

\begin{figure}
\includegraphics[width=9 cm]{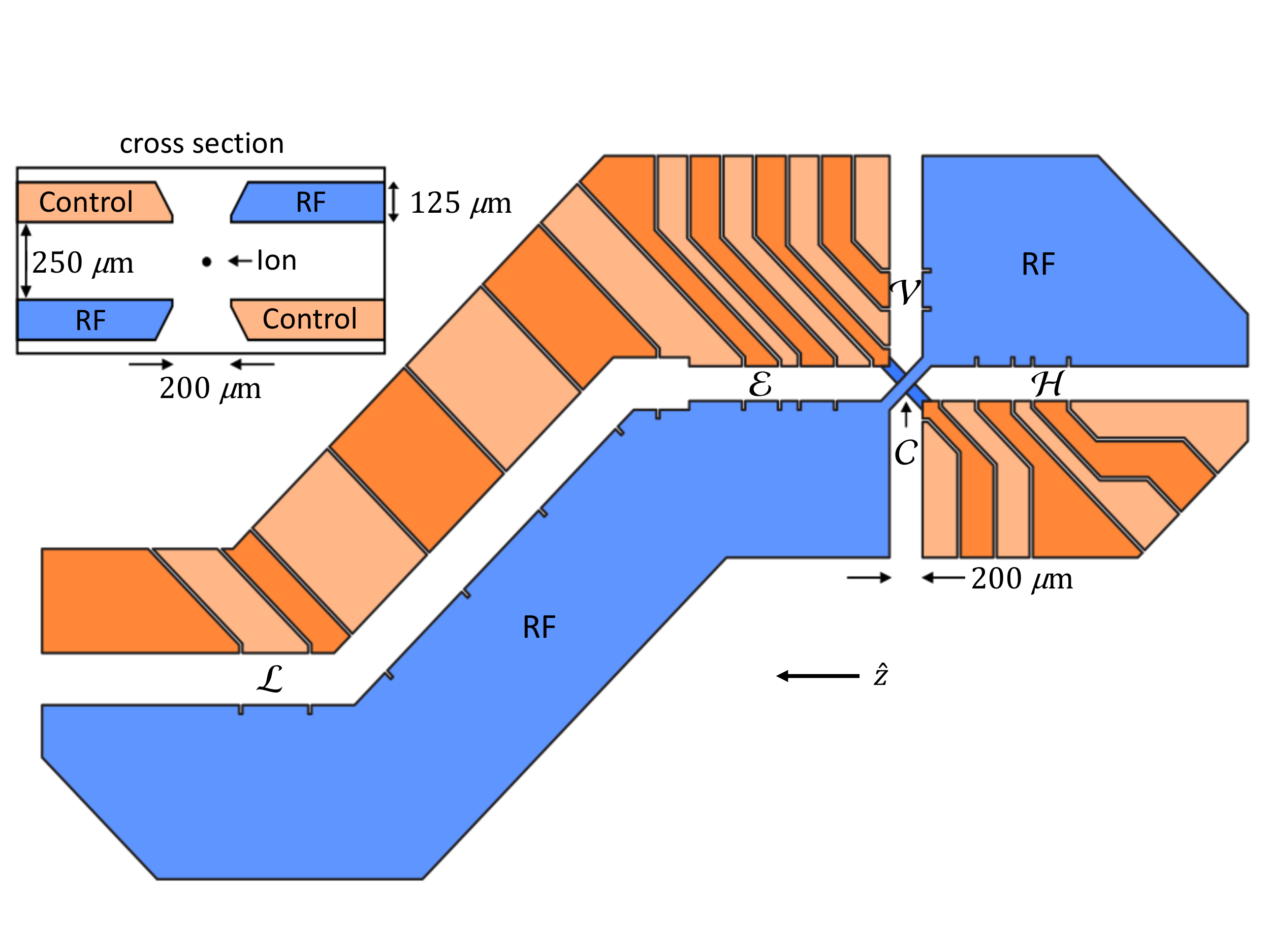}
\caption{Schematic of the ion trap, formed with two gold-coated, stacked wafers. Top view of the trap (on the right) showing the load zone $\mathcal{L}$ and experiment zone $\mathcal{E}$. Ions are transported from $\mathcal{L}$ to $\mathcal{E}$ with time-varying potentials applied to the segmented control electrodes (colored orange hues). The positions of RF and control electrodes are exchanged in the lower layer (cross section in inset). Coherent manipulations are implemented on ions confined in $\mathcal{E}$. Reprinted from \cite{Blakestad2011}.}
\label{fig:Xtrap}
\end{figure}

The ions are confined in a multi-segmented linear Paul trap (Fig.\ref{fig:Xtrap}) designed to demonstrate scalable QIP  \cite{Blakestad2011, bible,Kielpinski2002}. Radio frequency (RF) potentials, with frequency $\omega_{\mathrm{RF}} \simeq 2\pi \times 83$ MHz and amplitude $V_{\mathrm{RF}} \simeq $ 200 V, are applied to the RF electrodes to provide confinement transverse to the main trap channels. DC potentials are applied to the segmented control electrodes to create potential wells for trapping of ions at desired locations in the channels. By applying time-dependent potentials to these electrodes, the ions can be transported deterministically between different trap zones. The trap also contains a junction at $\mathcal{C}$, which can be used for reordering \cite{Blakestad2011}. For the experiment here, the ions are first loaded in $\mathcal{L}$ and then transported to $\mathcal{E}$. Quantum logic experiments described below are performed with ions confined in a fixed harmonic well at $\mathcal{E}$. Due to the particular design of the junction and trap imperfections, the ions undergo residual RF  ``micromotion" at frequency $\omega_{\mathrm{RF}}$ along $\hat{z}$ with amplitude $\simeq$ 105 nm at $\mathcal{E}$. This affects our implementation of logic gates, Doppler and ground state cooling, and qubit state measurement, as described below. 

\begin{figure}
\includegraphics[width=1\linewidth]{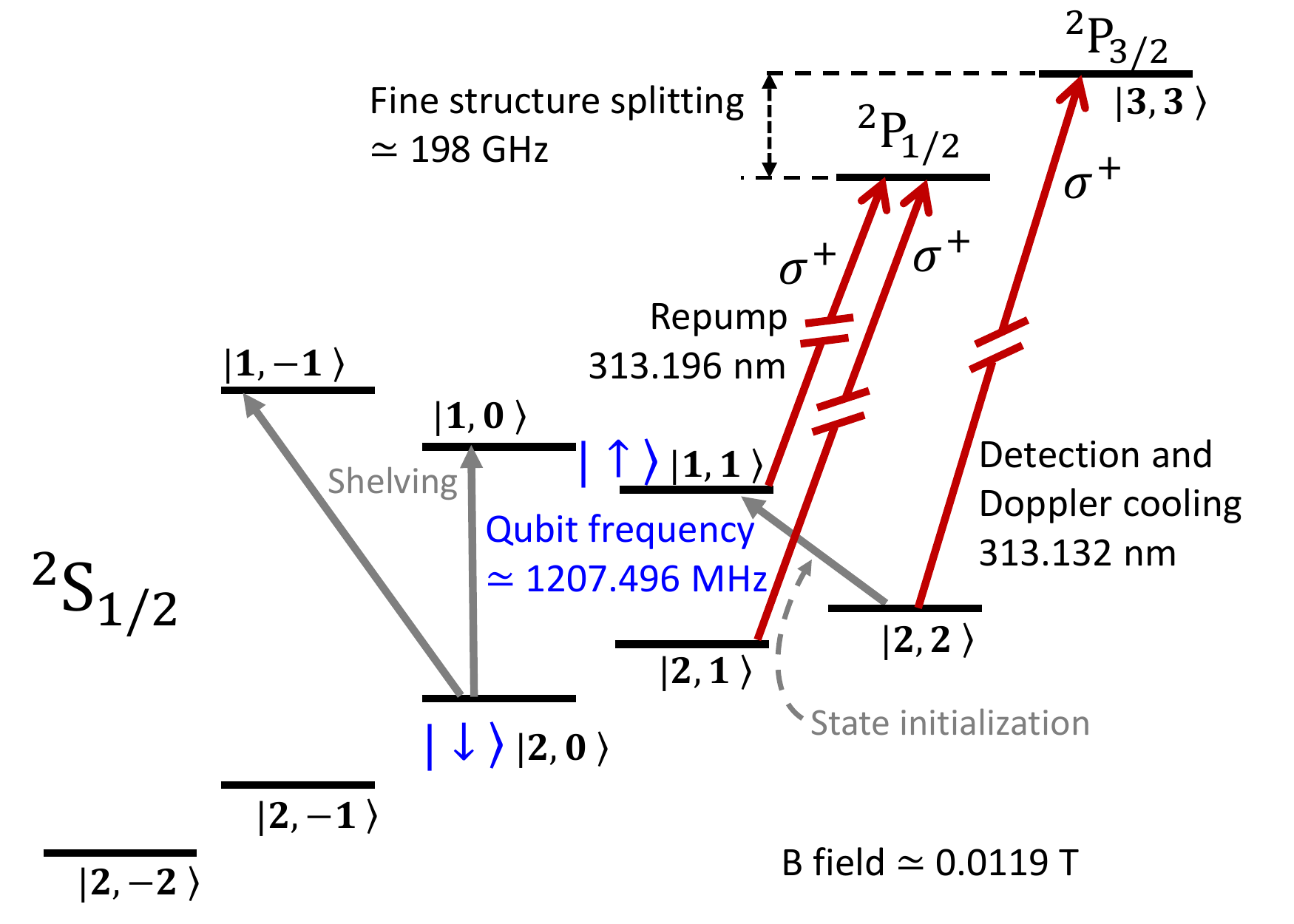}
\caption{Relevant energy level structure for $^{9}$Be$^+$ ions (not to scale).  Transitions to the electronic excited states are used for Doppler cooling, repumping, and qubit state measurement as described in the text.}
\label{fig:BeLevels}
\end{figure}

\begin{figure}
\includegraphics[width=1\linewidth]{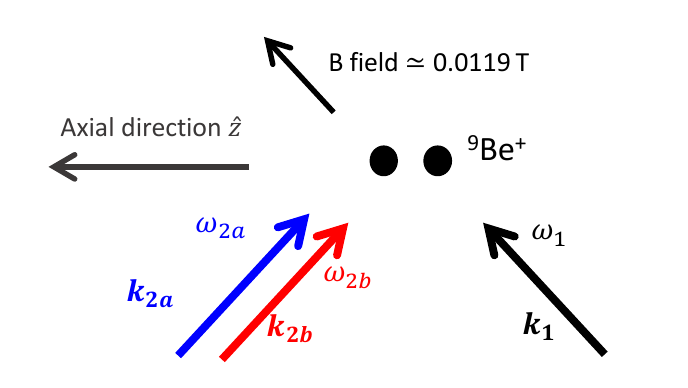}
\caption{Laser beam geometry for stimulated-Raman transitions. Co-propagating beams 2a and 2b are used to implement high-fidelity single qubit gates; two-qubit entangling gates use all three beams as described in the text.}
\label{fig:BeamLineSetup}
\end{figure}

For a single $^9$Be$^+$ ion confined in $\mathcal{E}$, the axial $z$ harmonic mode frequency is  $\omega_z \simeq 2 \pi\times 3.58$ MHz, while the transverse mode frequencies are $\omega_x \simeq 2 \pi\times 11.2$ MHz, and  $\omega_y \simeq 2 \pi\times 12.5$ MHz.   The ground state hyperfine levels and relevant optical levels for $^{9}$Be$^+$ ions in a magnetic field B $\simeq$ 0.0119 T are shown schematically in Fig. \ref{fig:BeLevels}. The qubit is encoded in the $^2$S$_{1/2}\ket{F=2, m_F = 0}  = \ket{\downarrow}$ and $\ket{1,1} = \ket{\uparrow}$ hyperfine levels, where $F$ and $m_F$ are total angular momentum and its projection along the quantization axis, respectively. The qubit frequency, $\omega_0 = 2\pi \times f_0 \simeq 2\pi \times 1207.496$ MHz is first-order insensitive to magnetic field fluctuations \cite{Langer2005}; we measure a coherence time of approximately 1.5 s. Before each experiment, we Doppler cool and optically pump the ion(s) to the $\ket{2,2}$ state with three laser beams that are $\sigma^+$-polarized relative to the B field and drive the $^2$S$_{1/2}\ket{2,2} \rightarrow ^{2}$P$_{3/2}\ket{3,3}$ cycling transition as well as deplete the $\ket{1,1}$ and $\ket{2,1}$ states (Fig. \ref{fig:BeLevels} and supplementary material). Both ions are then initialized to their $\ket{\uparrow}$ state by applying a composite $\pi$ pulse on the $\ket{2,2} \rightarrow \ket{\uparrow}$ transition.  After gate operations and prior to qubit state detection, population in the $\ket{\downarrow}$ state is transferred or ``shelved" to either the $\ket{1,-1}$ or $\ket{1,0}$ state and the $\ket{\uparrow}$ state is transferred back to the $\ket{2,2}$ state (supplementary material). We then apply the Doppler-cooling beam and observe fluorescence. In the two-ion experiments, for a detection duration of 330 $\mu$s, we detect on average approximately 30 photons for each ion in the $\ket{\uparrow}$ state, and approximately 2 photons when both ions are in the $\ket{\downarrow}$ state.  Coherent qubit manipulation is realized via two-photon stimulated-Raman transitions \cite{Monroe1995,bible} (supplementary material).  The required laser beams (Fig. \ref{fig:BeamLineSetup}) are directed to the trap via optical fibers \cite{Colombe2014} and focused to beam waists of approximately 25 $\mu$m at the position of the ions. 

High-fidelity single-qubit gates are driven with co-propagating beams $\bm{k_{2a}}$ and $\bm{k_{2b}}$ detuned by $\Delta$ from the $^2$S$_{1/2} \leftrightarrow ^2$P$_{1/2}$ transition frequency with their frequency difference set to $\omega_0$.  In this co-propagating beam geometry, single-qubit gates are negligibly affected by ion motion.  We employ the randomized benchmarking technique described in \cite{Brown2011} to characterize gate performance. Each computational gate consists of a Pauli gate ($\pi$ pulse) followed by a (non-Pauli) Clifford gate ($\pi/2$ pulse) around the $x$, $y$, and $z$ axes of the Bloch sphere, and identity gates. The $\pi$ pulses are performed with two sequential $\pi/2$ pulses about the same axis, each with duration $\simeq 2\ \mu$s. Rotations about the $z$ axis are accomplished by shifting the phase of the direct digital synthesizer that is keeping track of the qubit's phase; the identity gate is implemented with a $1\ \mu$s wait time. We deduce an error per computational gate of $3.8(1) \times 10^{-5}$. For $\Delta \simeq - 2 \pi \times$ 730 GHz used here, spontaneous emission error \cite{Ozeri2007} is estimated to be $2.5 \times 10^{-5}$.  The remaining error is dominated by Rabi rate fluctuations of approximately $1\times 10^{-3}$ due to imperfect laser power stabilization.

\begin{figure}
\includegraphics[width=1\linewidth]{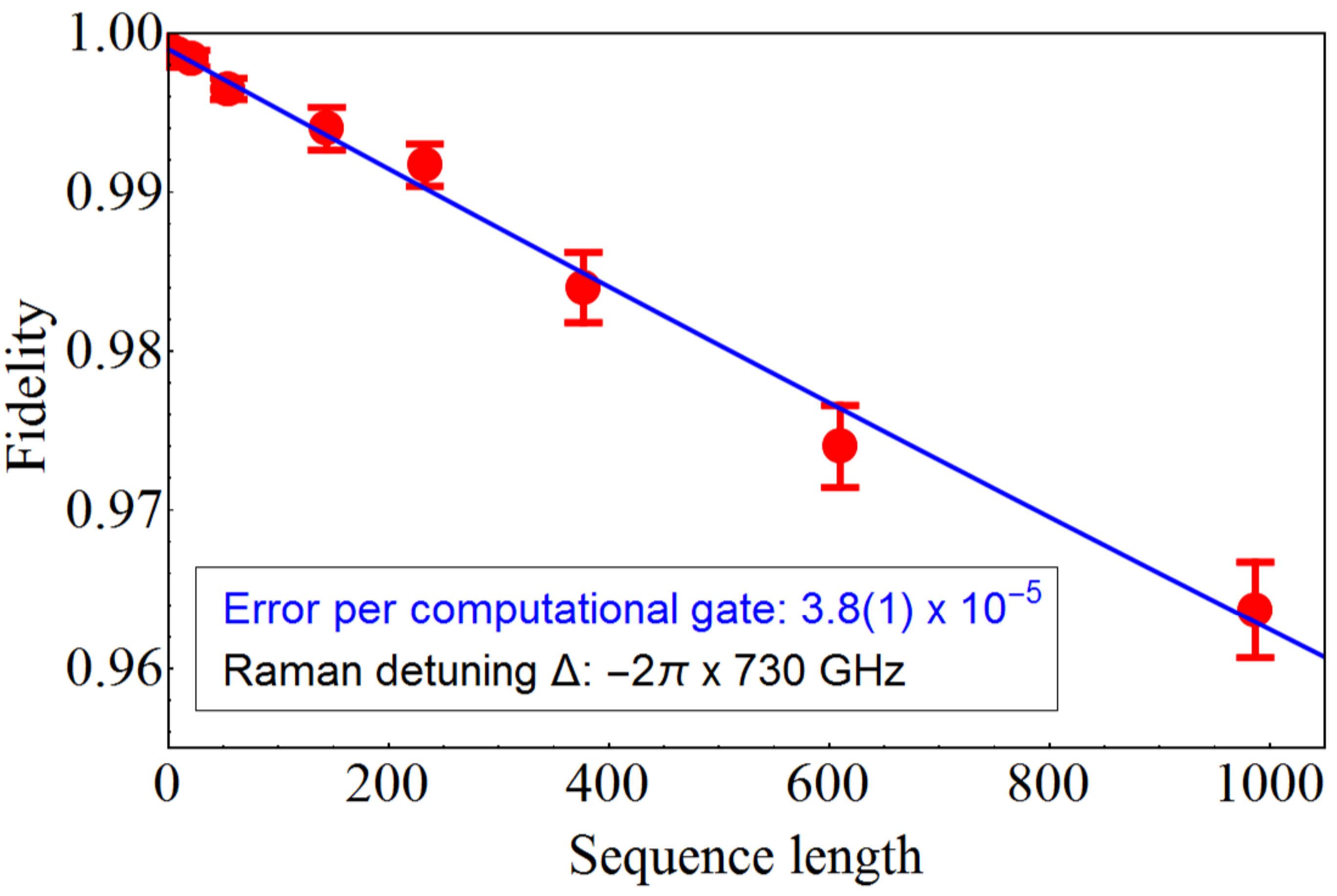}
\caption{Average fidelity for single-qubit-gate randomized benchmarking sequences, plotted as a function of sequence length. We determine the average error per computational gate to be $3.8(1)\times10^{-5}$ and state preparation and measurement error to be $2.0(3) \times 10^{-3}$ for these data sets. Error bars show the standard error of the mean for each point.}
\label{fig:RB}
\end{figure}

To couple the ions' internal (``spin") states to their motion, Raman transitions are driven by two beams along paths 1 and 2 respectively (Fig. \ref{fig:BeamLineSetup}).  These beams intersect at $90^{\circ}$ such that the difference in their $\bm{k}$ vectors, $\bm{\Delta k}$, is aligned along the axial direction, in which case only the axial motion will couple to the spins \cite{bible,Monroe1995}. The strength of the spin-motion coupling provided by these beams is proportional to the single-ion Lamb-Dicke parameter $\eta = |\bm{\Delta k}| z_{0} \simeq 0.25$ where $z_0 = \sqrt{\hbar/(2 m \omega_z)}$, with $\hbar$ and $m$ the reduced Planck's constant and the ion mass. However, due to the micro-motion along the axial direction, the carrier and spin-motion sideband Rabi rates are reduced for this laser beam geometry. For our parameters, the modulation index due to the micro-motion Doppler shift is approximately 2.9 such that the largest Rabi rates are provided by the second micro-motion sideband which is reduced by a factor of $J_2(2.9) \simeq 0.48$ relative to Rabi rates in the absence of micromotion.

Two trapped ions confined in $\mathcal{E}$ align along the axial direction with spacing 3.94 $\mu$m.  The relevant axial modes are the center-of-mass (C) mode (ions oscillate in phase at $\omega_z$) and and ``stretch" (S) mode (ions oscillate out of phase at $\sqrt{3} \omega_z$).  The two-qubit entangling gate is implemented by applying an effective $\hat{\sigma}_x\hat{\sigma}_x$ type spin-spin interaction using state-dependent forces (here acting on the axial stretch mode) in a M$\o$lmer-S$\o$rensen (MS) protocol \cite{Sorensen1999,Sorensen2000,Milburn1999,Solano1999} using all three beams in Fig. \ref{fig:BeamLineSetup} (supplementary material). To maximize the spin-motion coupling and state-dependent forces with the ions undergoing micromotion, the three beam frequencies are set to $\omega_1 = \omega_L$, $\omega_{2a} = \omega_L + 2\omega_{\mathrm{RF}} - \omega_0 + \omega_{\mathrm{S}} + \delta$, and $\omega_{2b} = \omega_L + 2\omega_{\mathrm{RF}} - \omega_0 - \omega_{\mathrm{S}} - \delta$, where $\omega_L$ is the laser frequency, which is detuned by $\Delta$ from the $^2$S$_{1/2} \rightarrow\ ^2$P$_{1/2}$ transition frequency, and $\delta$ is a small detuning ($\ll \omega_z$) that determines the gate duration \cite{Sorensen2000}.  Following initial Doppler cooling, the ions are sideband cooled with a series of $\ket{2,2}\ket{n} \rightarrow \ket{\uparrow}\ket{n-1}$ transitions, followed by repumping \cite{Monroe1995}, resulting in mean mode occupation numbers $\langle n_{\mathrm{C}} \rangle \simeq 0.01$ and $\langle n_{\mathrm{S}} \rangle \simeq 0.006$ and the ions being pumped to the $\ket{2,2}$ state. Two-qubit measurements are made as in the one ion case, but we collect fluorescence from both ions simultaneously.  We record photon count histograms with repeated experiments having the same parameters to extract the information about the qubit states.

\begin{figure}
\includegraphics[width=1\linewidth]{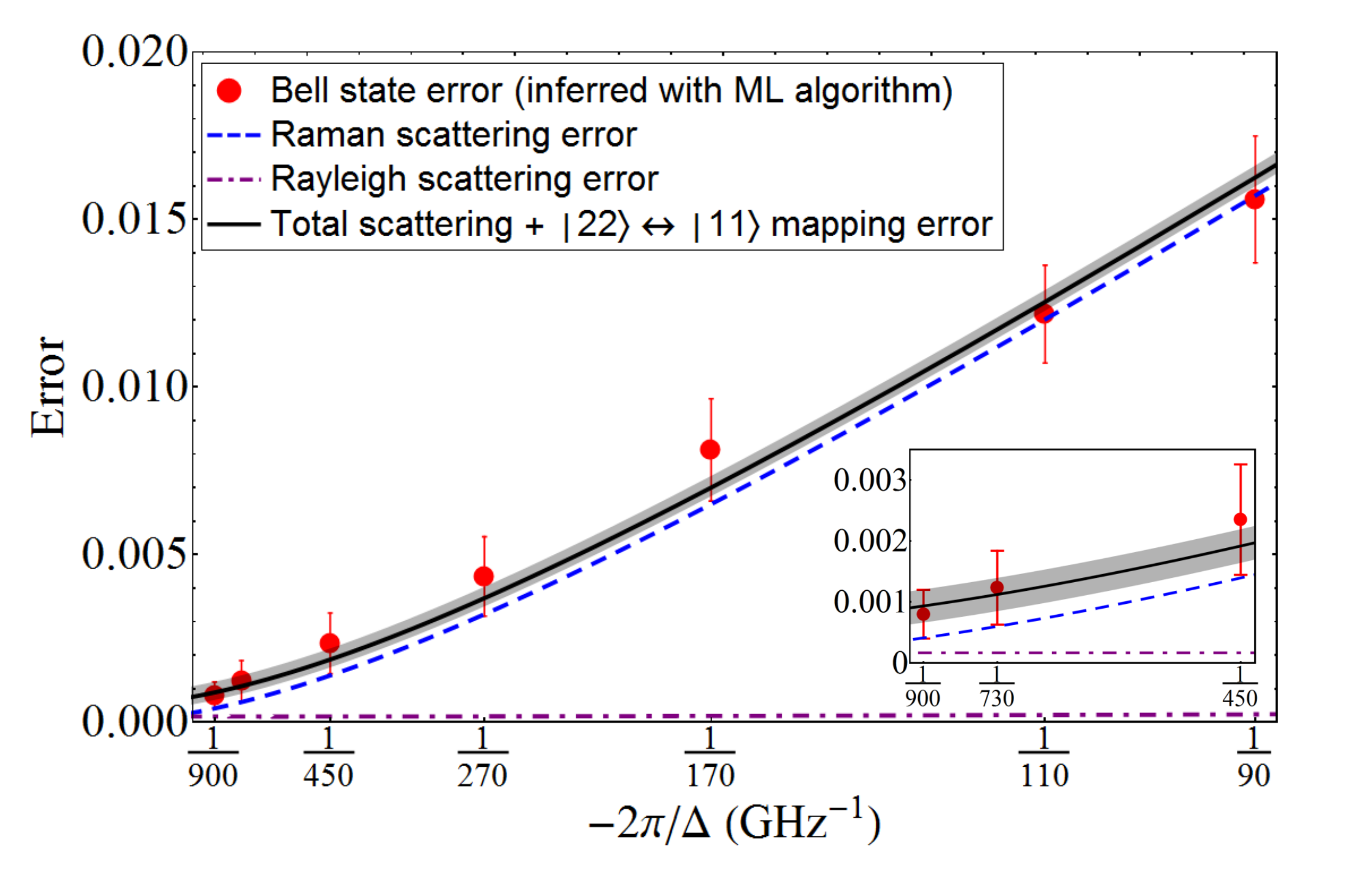}
\caption{ML-Bell-state error (red circles), plotted as a function of $-2 \pi/\Delta$ where $\Delta$ is the Raman detuning, for a constant gate duration of approximately 30 $\mu$s.  The simulated contributions to the Bell state error from Raman and Rayleigh scattering (supplementary material) are shown with the blue and purple dashed lines respectively. For large $|\Delta|$ the Raman scattering error approaches zero, however, the Rayleigh scattering error remains approximately constant at $1.7\times10^{-4}$.  The black line is the sum of the Raman and Rayleigh scattering errors and the composite microwave pulses used for qubit state preparation and detection (uncertainty indicated by the gray band). Error bars for the measured Bell state fidelity are determined from parametric bootstrap resampling \cite{Efron1993} of the data and represent a 1-$\sigma$ statistical confidence interval.}
\label{fig:GateDetuning}
\end{figure}

We use the gate to ideally prepare the Bell state $\ket{\Phi_+} = \textstyle{\frac{1}{\sqrt{2}}(\ket{\uparrow\uparrow}+\ket{\downarrow\downarrow})}$. To evaluate the gate's performance, we employ partial state tomography analyzed with a maximum likelihood (ML) algorithm to deduce the fidelity of the experimentally prepared state. Using a set of reference histograms, the maximum likelihood method estimates the experimentally created density matrix by maximizing the probability of the data histograms to correspond to that density matrix. The ML algorithm is general enough that joint-count histograms (here photon counts from two ions) can be analyzed without the need for individual addressing and measurement. From the Bell-state fidelity as determined by the ML method, we can estimate the MS gate fidelity.  The ML-Bell-state fidelity does not include errors due to imperfect $\ket{2,2}$ state preparation and measurement.  By taking these effects into account we also determine a lower bound for the actual Bell-state fidelity (supplementary material). 

\begin{figure}
\includegraphics[width=1\linewidth]{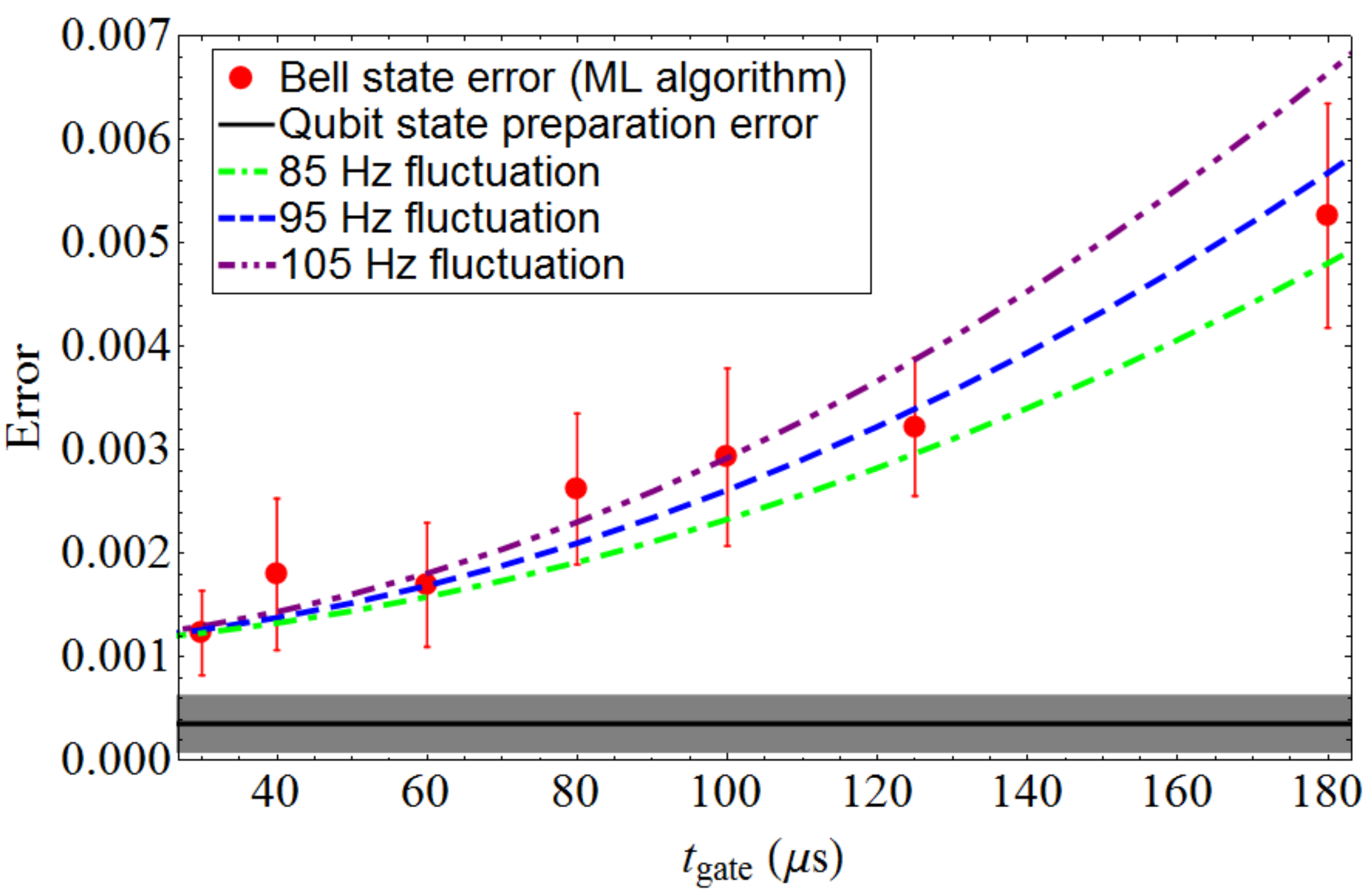}
\caption{ML-Bell-state error (red circles) as a function of gate duration $t_{\mathrm{gate}}$ for a constant Raman beam detuning  $\Delta \simeq - 2\pi \times 730$ GHz. The black line shows the separately determined error and uncertainty (gray shade) due to the microwave pulses used for $\ket{2,2} \leftrightarrow \ket{\uparrow}$ state transfer.  The three dashed lines show the sum of the expected gate errors including photon scattering and mode frequency fluctuations (which are slow compared to gate durations shown) for three different r.m.s. magnitudes of mode frequency fluctuations (supplementary material). The gate error increases quadratically with increasing $t_{\mathrm{gate}}$ due to such frequency fluctuations; however,  for $t_{\mathrm{gate}} = 30\ \mu$s the error due to frequency fluctuations is approximately $1 \times 10^{-4}$.}
\label{fig:GateDuration}
\end{figure}

By varying the laser beam power, we determine the error of the Bell state as a function of $\Delta$ keeping a fixed gate duration of $\simeq$ 30 $\mu$s (Fig. \ref{fig:GateDetuning}) and also as a function of gate duration for a fixed detuning $\Delta \simeq -2\pi \times 730$ GHz (Fig. \ref{fig:GateDuration}). The various curves in the figures show the expected errors due to spontaneous emission and errors in the composite microwave pulses used for $\ket{2,2} \leftrightarrow \ket{1,1} = \ket{\uparrow}$ state transfer, and mode frequency fluctuations in Fig. \ref{fig:GateDuration}. The minimum error obtained is $8(4) \times 10^{-4}$ for $\Delta \simeq - 2\pi \times 900$ GHz and a gate duration of approximately 30 $\mu$s, which yields a ML-Bell-state fidelity of 0.9992(4). An important contribution to the ML-Bell-state error is due to the imperfect transfers from the $\ket{2,2}$ state to the qubit $\ket{\uparrow}$ state (for both qubits) before the application of the gate, and the reverse procedure that transfers $\ket{\uparrow}$ population back to the $\ket{2,2}$ state before detection. The total fidelity of these transfer pulses, limited by magnetic field fluctuations and the quality of the microwave pulses, is investigated with separate experiments analyzed with the same ML algorithm (supplement), and we find $\epsilon_{\mathrm{transfer}} = 4(3)\times 10^{-4}$. This is averaged over multiple data evaluations across multiple days; the uncertainty is the standard deviation of these data. While this error does not in principle affect the gate performance, we conservatively do not remove it from our gate fidelity estimate due to its relatively large uncertainty.

In the supplementary material, we describe in more detail characterization of individual errors sources through calibration measurements and numerical simulation.  From this, we deduce that the fidelity of the ML-Bell-state is a good representation of the average gate fidelity. The errors for the highest state fidelity obtained are listed in Table \ref{ErrorBudget}.  It would be advantageous to evaluate the gate performance with full process tomograghy or randomized benchmarking to confirm our assessment. We did not perform randomized benchmarking because ion motional excitation gives additional errors.  This excitation occurs during ion separation (to provide individual ion addressing) and because of anomalous heating \cite{Turchette2000} during the required long sequences of gates. These problems can eventually be solved as in \cite{Gaebler2012} where the gate fidelity was measured by interleaved randomized benchmarking or by process tomography \cite{Navon2014}. In both cases, the gate error was consistent with the measured two-qubit state fidelity. In the experiment here, the uncertainties of the inferred errors are deduced by parametric bootstrap resampling \cite{Efron1993} with 500 resamples. We determine a lower bound of $0.999$ on the purity of the $\ket{2,2}$ state for one ion prepared by optical pumping. With this, we put a lower bound of $0.997$ on the overall Bell state fidelity. 

\begin{table}
\begin{tabular}{|c|c|c|}
  \hline
  Errors & $\times 10^{-4}$\\
  \hline
  Spontaneous emission (Raman) & $4.0$\\
	Spontaneous emission (Rayleigh) & $1.7$\\
  Motional mode frequency fluctuations & 1\\ 
	Rabi rate fluctuations & 1\\
	Laser coherence & 0.2 \\
	Qubit coherence & $<$0.1 \\
  Stretch-mode heating & $ 0.3$\\
	Error from Lamb-Dicke approximation & 0.2\\
	Off-resonant coupling & $<$0.1 \\
	\hline
	$\ket{2,2} \Leftrightarrow \ket{\uparrow}$ two-way transfer & 4\\
  \hline
\end{tabular}
\caption{Error budget for the entangling gate at a Raman detuning of $\Delta \simeq - 2\pi \times 900$ GHz, and a gate duration of 30 $\mu$s. Off-resonant coupling includes coupling of the qubit states to other hyperfine states and their sidebands.  The last error reduces the ML-Bell-state fidelity but should minimally affect the gate fidelity.}
\label{ErrorBudget}
\end{table}

In summary, we have demonstrated high fidelity single- and two-qubit laser-induced gates on trapped $^9$Be$^+$ ions. The single-qubit gate fidelity exceeds some threshold estimates for fault-tolerant error correction with reasonable overhead.  Sources of the $\simeq 10^{-3}$ two-qubit gate error have been identified and can likely be reduced, making $^9$Be$^+$ ion a strong qubit candidate for fault-tolerant QIP. Gates with comparable fidelity have been recently reported by the Oxford group using $^{43}$Ca$^+$ ions \cite{Ballance2015}.

This work was supported by the Office of the Director of National Intelligence (ODNI) Intelligence Advanced Research Projects Activity (IARPA), ONR and the NIST Quantum Information Program. We thank D. Allcock and S. Brewer for helpful suggestions on the manuscript. We thank D. Hume, D. Lucas, C. Ballance and T. Harty for helpful discussions. This paper is a contribution of NIST and not subject to U.S. copyright.

\section{Supplementary Material}

\subsection{Laser beam configuration}

Four infrared lasers are used to create three separate UV sources: 313.132 nm for the $^2$S$_{1/2}$ to $^2$P$_{3/2}$ transitions, 313.196 nm for the $^2$S$_{1/2}$ to $^2$P$_{1/2}$ transitions, and a variable wavelength source in the range of approximately 313.260 to 313.491 nm for Raman transitions. We generate visible light near 626 nm by employing sum frequency generation (SFG) of a pair of infrared laser beams (one near 1050 nm and the other near 1550 nm) using  temperature-tuned magnesium-oxide-doped periodically-poled lithium-niobate (MgO:PPLN) in a single-pass configuration. These lasers setups are similar to that described in \cite{Wilson2011} and provide up to 2.5 W in each visible beam with relatively low power fluctuations ($<1\ \%$ rms). The visible light is then frequency doubled using a Brewster-angle cut beta barium borate (BBO) crystal in a bowtie cavity configuration \cite{Wilson2011}. A separate mode-locked pulsed laser system near 235 nm is used for photo-ionization of neutral $^9$Be atoms during loading of ions into the trap. 

The $^2$S$_{1/2}$ to $^2$P$_{3/2}$ beam is used for Doppler cooling and qubit state measurement. Doppler cooling is achieved with $\sigma^+$ polarized light red detuned by approximately $\Gamma/2$ (angular frequency) from the $^2$S$_{1/2}\ket{2,2}$ to $^2$P$_{3/2}\ket{3,3}$ transition at 313.132 nm (Fig. \ref{fig:BeLevels}), where $\Gamma \simeq 2\pi \times 19.6$ MHz is the decay rate of the $^2$P$_{3/2}$ state. The Doppler cooling beam $\bm{k}$-vector direction is such that it can cool all modes of the ions' motion. During Doppler cooling and detection, to maximize efficiency in the presence of the axial micromotion, we apply a differential voltage of approximately $\pm\ 0.15$ V to the two control electrodes centered on zone $\mathcal{E}$. This shifts the ions away from the radial micromotion null point (trap axis) such that the vector sum of the radial and axial micromotion is perpendicular to the Doppler cooling beam's wavevector. Qubit state-dependent fluorescence detection is accomplished by first reversing the initial qubit state preparation to transfer the $\ket{\uparrow}$ population back to the $\ket{2,2}$ state.  This is followed by a microwave $\pi$ pulse that transfers (``shelves") the $\ket{\downarrow}$ state to the $\ket{1,-1}$ state. To shelve any remaining $\ket{\downarrow}$ population, we then apply a microwave $\pi$ pulse from the $\ket{\downarrow}$ state to the $\ket{1,0}$ state. After these shelving pulses, the $^2$S$_{1/2}$ to $^2$P$_{3/2}$ laser beam is tuned to resonance on the $^2$S$_{1/2}\ket{2,2}$ to $^2$P$_{3/2}\ket{3,3}$ cycling transition. With these conditions, the fluorescing or ``bright'' state of this protocol corresponds to the qubit $\ket{\uparrow}$ state, and the qubit $\ket{\downarrow}$ state will be detected as ``dark''. With a detection duration of $330\ \mu$s, we record on average approximately 30 photon counts in a photo-multiplier tube for an ion in the bright state and 2 photons for 2 ions in the dark state (limited by background scattered light). 

The $^2$S$_{1/2}$ to $^2$P$_{3/2}$ Doppler cooling and detection laser beam will optically pump the ions to the  $^2$S$_{1/2}\ket{2, 2}$ state as long as the beam has pure $\sigma^+$ polarization with respect to the B field.  To mitigate the effects of polarization impurity and speed up the pumping process, two $^2$S$_{1/2}$ to $^2$P$_{1/2}$ laser beams are added for the initial optical pumping to the $^2$S$_{1/2}\ket{2, 2}$ state. All three beams are first applied, and the final stage of pumping uses only the  $^2$S$_{1/2}$ to $^2$P$_{1/2}$ beams. These beams are derived from the same laser source that is split into two, with one beam frequency tuned near the $^2$S$_{1/2}\ket{2, 1}$ to $^2$P$_{1/2}\ket{2, 2}$ transition, and the other tuned near the $^2$S$_{1/2}\ket{1, 1}$ to $^2$P$_{1/2}\ket{2, 2}$ transition. To suppress electromagnetically-induced transparency effects that would lead to coherent trapping of population in the $^2$S$_{1/2}\ket{2, 1}$ and the $^2$S$_{1/2}\ket{1, 1}$ states when the beams are applied simultaneously and on resonance, one of these beams is detuned from its atomic resonance by approximately $-\Gamma/2$. These beams also serve to repump to the $^2$S$_{1/2}\ket{2, 2}$ state during Raman sideband cooling \cite{Monroe1995}. All of these UV beams are then overlapped inside a UV optical fiber \cite{Colombe2014} before being focused onto the ions' location. 

\begin{figure}
\includegraphics[width=1\linewidth]{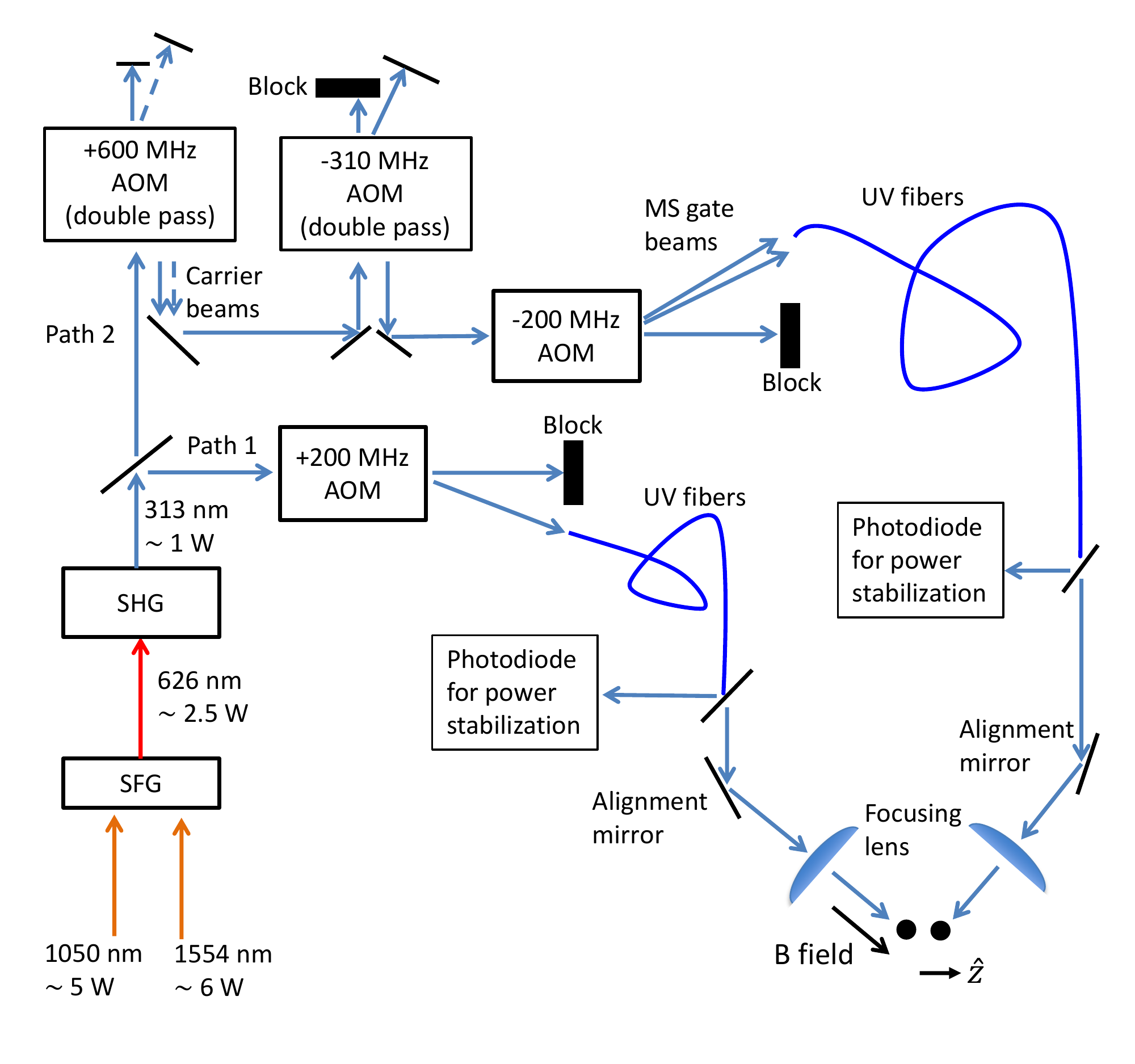}
\caption{A schematic of the laser setup used for stimulated-Raman laser gates.  The 313 nm light is generated from two IR sources by sum frequency generation (SFG) followed by second harmonic generation (SHG).  The beam is split into two paths, sent through AOMs, coupled into fibers, and aligned onto the ions. Path 2 contains a double-passed 600 MHz AOM that, when switched on, produces an additional beam shifted by approximately the qubit frequency, $f_{0}$, that is co-aligned with the unshifted beam for high-fidelity single-qubit gates. Another AOM tunable in the range of 260 to 360 MHz is used to shift the relative laser frequency in path 2 with respect to beam 1 for Raman sideband transitions. For the M$\o$lmer-S$\o$rensen gate, two RF tones with relative frequency difference close to twice the frequency of the addressed motional mode are injected into the 200 MHz AOM in path 2.  In combination with the beam in path 1, these two beams simultaneously produce blue and red sideband transitions. A pickoff on the output of each fiber directs a small fraction of the light onto a photodiode, which is used for active power stabilization. Each beam is centered on the ions with a motorized mirror mount before the final lens.}
\label{fig:LaserSetup}
\end{figure}

The laser beams for qubit manipulation with stimulated-Raman transitions ($\lambda \simeq$ 313 nm) are detuned by $\Delta$ from the $^2$S$_{1/2}$ to $^2$P$_{1/2}$ electronic transitions. The UV beam is first split and sent down two different paths (Fig. \ref{fig:LaserSetup}). The beam in path 1 passes through an acousto-optic modulator (AOM) where the first-order deflected beam (+ 200 MHz) is coupled into an optical fiber. The output beam from the optical fiber is then focused to a waist of $\simeq 25\ \mu$m at the location of the ions. The beam in path 2 is first sent through a double-pass AOM with center a frequency of 600 MHz. The 600 MHz AOM geometry is configured such that when it is switched off, it simply outputs the input beam.  When it is switched on, it outputs an additional beam shifted by $\simeq +\ 2\times 600$ MHz that is co-alligned with the unshifted beam. In both cases, the output of the 600 MHz AOM is then sent through a double-pass AOM with a center frequency of 310 MHz followed by a single pass 200 MHz AOM in a setup analogous to that in path 1.  The tuning range between beams in paths 1 and 2 is approximately 200 MHz. The frequency and phase of the RF for each AOM is generated with computer-controlled direct digital synthesizers (DDS) that are phase stable relative to each other.

The optical fibers are robust against color center formation \cite{Colombe2014} and substantially suppress higher-order modes compared to the case when they are not used.  (These fibers have been extensively used for over 18 months with output of powers up to 100 mW and no observed degradation in the transmission ($\simeq 50\ \%$) or mode quality. Also, we do not observe significant polarization drift.)  Furthermore, since the fiber outputs are located relatively close to the beam input windows of the vacuum chamber, beam intensity fluctuations at the ions' location due to air currents, vibrations, and temperature drifts are reduced compared to similar beam lines set up in free-space. Beam position fluctuations at the input to the fibers translate to laser power fluctuations at the fiber outputs. The laser power at each of the fiber's output is stabilized by monitoring a sample of the beam on a photodiode and feeding back to the RF power that drives the 200 MHz AOMs in each path. Each feedback loop is controlled by a Field Programmable Gate Array (FPGA) based digital servo \cite{Leibrandt2015}, while another FPGA-based digital-to-analog converter (DAC) \cite{Bowler2013} is used to actively control the servo set point that provides beam temporal pulse shaping to smooth the transitions between on and off. Turn on and off durations for the pulses are approximately 0.75 $\mu$s. 

For high-fidelity single-qubit gates, the 200 MHz AOM in path 1 is switched off and the 600 MHz AOM in path 2 is switched on. The resulting two beams in path 2 have a frequency difference that can be tuned to the qubit frequency, $f_0$. The qubit transition in this copropagating configuration is insensitive to the ion's motion, and we measure insignificant phase drift between these two laser beams during single experiment. The polarization of the these beams is set at $45^{\circ}$ with respect to $\bm{k_1}$ and the B-field direction (Fig. \ref{fig:BeamLineSetup}) such that they contain equal parts $\pi$ and $\sigma^{+,-}$ polarization at the ion's location.  

Stimulated-Raman sideband transitions that couple the spins (hyperfine states) to the motion are driven by laser beams from paths 1 and 2.  Since the second axial micromotion sideband provides the largest coupling strength, we tune the difference frequency of the Raman beams to near $\omega + 2\omega_{\mathrm{RF}}$ or $\omega - 2\omega_{\mathrm{RF}}$, where $\omega$ is the transition frequency of interest. Before applying the MS gate, the ions are first Doppler cooled followed by ground-state cooling on the axial modes \cite{Monroe1995}. Ground state cooling is accomplished by first driving a series of red-sideband pulses on the $\ket{2, 2}$ to $\ket{\uparrow}$ transition (which has carrier frequency $\omega_{\ket{2,2}\leftrightarrow \ket{1,1}} \simeq 2\pi \times 1018$ MHz), each followed by repumping with the $^2$S$_{1/2}$ to $^2$P$_{1/2}$ beams to re-initialize the ions back to the $\ket{2, 2}$ state. We first apply 10 cooling pulses on the second motional sideband of the center-of-mass (C) mode (Raman beam frequency difference set to $\omega_{\ket{2,2}\leftrightarrow \ket{1,1}} + 2\omega_{\mathrm{RF}} - 2 \omega_{\mathrm{C}})$  followed by 60 pulses each on the first motional sidebands (at $\omega_{\ket{2,2}\leftrightarrow \ket{1,1}} + 2\omega_{\mathrm{RF}} - \omega_{\mathrm{C,S}})$. These red sideband transitions are alternately applied on the C and stretch (S) modes.  With this, we achieve final mean occupation numbers, $\langle n \rangle \simeq$ 0.01 and 0.006 respectively \cite{Monroe1995}. The ions are then transferred from the $\ket{2,2}$ to the qubit $\ket{\uparrow}$ state by applying a composite microwave $\pi$ pulse \cite{Levitt1986} composed of a sequence of ($\theta,\phi)$ pulses $(\pi,0),\ (\pi,\pi/3),\ (\pi,\pi/6),\ (\pi,\pi/3),\ (\pi,0)$, where $\theta$ denotes the angle the state is rotated about an axis in the {\it x-y} plane of the Bloch sphere, and $\phi$ is the azimuthal angle of the rotation axis.

\subsection{M$\o$lmer-S$\o$rensen Gate}

Two trapped ions are aligned along the axial $z$ direction with spacing of $\simeq 3.94$ $\mu$m.  Their $z$ motion can be described by two normal modes, the center-of-mass (C) and stretch (S) modes with frequencies $\omega_{\mathrm{C}} =\omega_z$ and $\omega_{\mathrm{S}} = \sqrt{3}\omega_z$ respectively. The motion of the $i$th ion is written $z_i = z_{i,\mathrm{C}0}(a + a^{\dag}) + z_{i,\mathrm{S}0} (b + b^{\dag})$ where $a, a^{\dag} $ and $b, b^{\dag}$ are the lowering and raising operators for the C and S modes and $z_{1,\mathrm{C}0} = z_{2,\mathrm{C}0} = z_0/\sqrt{2}, z_{1,\mathrm{S}0} = - z_{2,\mathrm{S}0} = z_0/\sqrt{2\sqrt{3}}$.  The M$\o$lmer-S$\o$rensen (MS) interaction requires simultaneously driving a blue sideband with a detuning of $\delta$ and a red sideband with a detuning of $-\delta$ on the selected (stretch) mode. The sideband transitions are driven on the second order micromotion sideband with three laser beams as described in the main text. The difference in diffraction angle of beams $k_{2a}$ and $k_{2b}$ is small enough that they can both be coupled into the same single-mode fiber by imaging the center of the AOM into the fiber. The near perfect overlap of these two beams ensures any optical path fluctuations leading to laser beam phase fluctuations will be common and the stability of their phase difference is determined by the stable RF oscillators used to create the two tones. The third beam is generated in path 1; its frequency differs from the mean of the path 2 beams by $\omega_0 - 2 \omega_{\mathrm{RF}}$. The polarization of this beam is adjusted such that the power ratio of $\sigma^+$ to $\sigma^-$ components is $8:2$, with the $\sigma^+$ component used for driving the MS gate and the $\sigma^-$ component used for sideband cooling. 

Transforming to the interaction frames for both the spins and stretch mode of motion and dropping high frequency terms (rotating-wave approximation), the effective $\hat{\sigma}_{x} \hat{\sigma}_{x}$ M$\o$lmer-S$\o$rensen interaction in the Lamb-Dicke limit can be written as
\begin{align}
H = \hbar \sum_{j=1,2} \eta_{\mathrm{S}} \Omega \hat{\sigma}_j^+ \left(\hat{b} e^{-i(\delta t+\phi_{j,r})}+\hat{b}^{\dag} e^{i(\delta t-\phi_{j,b})}\right)+h.c.,
\label{EqMS}
\end{align}
where $\Omega$ is the resonant carrier transition Rabi rate, $\eta_{\mathrm{S}} = |\bm{\Delta k}| z_{1,\mathrm{S}0} \simeq 0.19$, $\hat{\sigma}_j^+$ is the spin raising operator for the $j$th ion, and $\phi_{j,b (r)}$ is the phase of the blue (red) sideband interaction on the $j$th ion. Starting in the $\ket{\uparrow\uparrow}$ state and setting $\eta_{\mathrm{S}} \Omega = \delta/2$, this interaction produces the state
\begin{align}
\frac{1}{\sqrt{2}}\left(\ket{\uparrow\uparrow} + 
e^{-i\left(\sum_{j=1,2}\frac{1}{2}(\phi_{j,b}+\phi_{j,r}) + \pi/2 \right)}\ket{\downarrow\downarrow}\right)
\label{EqPhiMS}
\end{align}
after a duration $2\pi/\delta$.  

From Eq. (\ref{EqPhiMS}), we see that the phase difference between the two components of the state depends on the phases $\phi_{j,b}$ and $\phi_{j,r}$, which can fluctuate over the course of repeated experiments due to relative path length fluctuations between paths 1 and 2 caused by air currents, mechanical vibrations, and thermal drifts. However, if all single-qubit pulses in an individual experimental sequence are applied using laser beams propagating along the same two beam paths, then the relative phases of the two-qubit gate and single qubit pulses will be stable as long as the beam path lengths are constant for the duration of each experiment. In this case we can choose the phase factors in Eq. (\ref{EqPhiMS}) for each experiment such that we realize the state $\ket{\Phi_+} = \textstyle{\frac{1}{\sqrt{2}}}(|\uparrow\uparrow\rangle + |\downarrow\downarrow\rangle)$.  Single-qubit gates driven by non-copropagating laser beams have the disadvantages of being sensitive to the ions' motion and will have lower relative phase stability compared to the case of co-propagating laser beams. However, by surrounding the MS gate pulse with two global $\pi/2$ single-qubit pulses (using the same non-co-propagating laser beams) a phase insensitive two-qubit gate that implements $\ket{\uparrow\uparrow} \rightarrow \ket{\uparrow\uparrow},\ \ket{\uparrow\downarrow} \rightarrow i\ket{\uparrow\downarrow},\ \ket{\downarrow\uparrow} \rightarrow i\ket{\downarrow\uparrow},\ \ket{\downarrow\downarrow} \rightarrow \ket{\downarrow\downarrow}$ can be achieved and all other single-qubit pulses can be performed with any phase stable source \cite{Lee2005,Tan2015}. 

\subsection{State detection and tomography}

For the single-qubit experiments, the photon-count histograms of the bright and dark states are well separated so we determine the ion's state by setting a threshold count, typically at 12 counts.  We estimate the state-detection error from this simple method to be $\sim 2 \times 10^{-3}$. Randomized benchmarking separates this error from the much smaller error per computational gate. For the two ion experiments, joint photon-count histograms are collected by recording state-dependent fluorescence counts from both ions with a detection laser beam size much larger than the ion separation. As a result, the recorded histograms are drawn from mixtures of three possible count distributions $q_j(c)$ ($c = 0,1,2,....,C$ indicates the photon counts) corresponding to the distinguishable ion subspaces spanned by (i) $\ket{\uparrow\uparrow}$, (ii) $\ket{\uparrow \downarrow}$ or $\ket{\downarrow \uparrow}$, and (iii) $\ket{\downarrow \downarrow}$ states. Because of the finite efficiency of our photon collection apparatus and optical pumping during detection, the three count distributions overlap, particularly those of subspaces (ii) and (iii).  Therefore an exact determination of the subspace to which the ions are projected cannot be determined in a single detection.  Nevertheless, we can infer the ions' density matrix statistically from repetitions of the experiment provided that the count distributions for each projected subspace are known.  These distributions can be inferred from reference experiments by fitting to a parametrized model of the distributions. A common class of such models is given by mixtures of Poissonians with different means. The uncertainty requirements of our experiments and effects such as optical pumping during photon collection imply that we cannot use such models unless they have an excessively large number of parameters, in which case overfitting becomes an issue.  Our maximum likelihood (ML) analysis avoids these issues by statistically inferring states without requiring a model for the ideal count distributions.

The ML analysis requires reference and data histograms, where the data histograms involve observations of an identically prepared state $\rho$ modified by analysis pulses. It infers a representative density matrix $\hat{\rho}$. Because the different observations are not ``informationally complete'', $\hat{\rho}$ is not intended to match $\rho$ {precisely but the measurements are designed so that the fidelities of interest do match to within a statistical uncertainty.  In our experiment, we obtain four reference histograms $r_i(c)$ ($i=1,2,3,4$). Each reference histogram is obtained by observing known ion states prepared as follows: For $r_1(c)$, the state is prepared by optical pumping both ions to the $\ket{2,2}$ state. For $r_2(c)$, this optical pumping is followed by implementing the transfer $\ket{2,2} \rightarrow \ket{\uparrow}$ with a composite microwave pulse, followed by the transfers $\ket{\uparrow}\rightarrow \ket{\downarrow}$ and shelving into one of the states $\ket{1,-1}$ or $\ket{1,0}$ with microwave $\pi$ pulses as described in the main text. For $r_3(c)$, the optical pumping is followed by the microwave-driven spin-echo sequence consisting of $(\frac{\pi}{2},0)$, $(\pi,0)$, $(\frac{\pi}{2},\frac{\pi}{2})$ pulses on the $\ket{2,2}$ to $\ket{\uparrow}$ transition, followed by transferring the population in the $\ket{\uparrow}$  state to the $\ket{1,-1}$ or $\ket{1,0}$ state as for $r_2(c)$. The histogram $r_4(c)$ is obtained like $r_3(c)$ but with the phase of the third pulse set to $\frac{3\pi}{2}$. The change in phase does not change the state when the initial state and pulses are as designed. Data histograms $h_k(c)$ are obtained directly from the prepared state $\rho$ or by applying analysis pulses on the prepared state. The analysis pulses are global $(\frac{\pi}{2},n\frac{\pi}{4})$ pulses, for $n = 0,1,...,7$. These pulses are applied using the laser beams from path 1 and 2 (Fig. \ref{fig:BeamLineSetup}) to maintain relative phase stability with respect to the two-qubit gate. 

To determine $\hat{\rho}$, we maximize the log(arithm of the) likelihood of the observed histograms with respect to the unknown $q_j(c)$ and $\rho$ to be determined. Given these unknowns, reference histogram $r_i(c)$ is sampled from the distribution $\sum_j a_{ij} q_j(c)$, where the $a_{ij}$ are ``populations'' determined from the known prepared state. Similarly, the data histograms $h_k(c)$ are sampled from the distribution $\sum_j b_{kj} q_j(c)$, where the populations $b_{kj}$ are a linear function of $\rho$. Given these distributions, the log likelihood is given by 

\begin{widetext}
\begin{align}
\mathrm{log}\left(\mathrm{Prob}(r,h|q,a,b)\right)=\left[\sum_{i=1,c}^{4,C}r_i(c)\mathrm{log}\left(\sum_{j=1}^3 a_{ij}q_j(c)\right)\right]+\left[\sum_{k,c}^{9,C} h_k(c)\mathrm{log}\left(\sum_j^3 b_{kj} q_j(c)\right)\right]+\mathrm{const}.
\end{align}
\end{widetext}
To maximize the log likelihood, we take advantage of the separate convexity of the optimization problem in the $q_j(c)$ and $\rho$ and alternate between optimizing with respect to the $q_j(c)$ and $\rho$. We used a generic optimization method for the first and the ``$R\rho R$'' algorithm \cite{Hradil2004} for the second to keep $\rho$ physical during the optimization. The quality of the model fit can be determined by a bootstrap likelihood-ratio test~\cite{Boos2003}. We found that the data's log-likelihood-ratio was within two standard deviations of the mean bootstrapped log-likelihood-ratio.

A refinement of the ML analysis is required to reduce the number of parameters needed for the $q_j(c)$ and the complexity of the algorithms. For this, we bin the counts into seven bins of consecutive counts \cite{Lin2016}. The binning is done separately by setting aside a random 10 $\%$ of each reference histogram and using this as a training set to determine a binning that maximizes state information quantified by a mutual-information-based heuristic. The heuristic is designed to characterize how well we can infer which of the four training set reference histograms a random count is sampled from. 

The ML analysis assumes that the reference histograms are sampled from count distributions corresponding to states with known populations. The actual populations deviate by small amounts from this assumption. We considered the systematic effects of such deviations on the analysis. Two effects were considered. One is that optical pumping that ideally prepares the $\ket{2,2}$ state may have fidelity as low as $0.999$ (see next section). The other is due to imperfections in the transfer pulses between the qubit manifold and other states.  This will be dominated by the transfer pulses between the $\ket{2,2}$ and $\ket{\uparrow}$ states, which has a measured fidelity of $0.9996(3)$. Errors in the transfer between the $\ket{\downarrow}$ and the $\ket{1,-1}$ and $\ket{1,0}$ states have less effect, since to a high degree all three of these states are dark.  To analyze the effects of these errors, we explicitly distinguish between the computational qubit based on the states $\ket{\uparrow}=\ket{1,1}$ and $\ket{\downarrow}=\ket{2,0}$ and the measurement qubit based on the bright state $\ket{2,2}$ and the dark states $\ket{1,-1}$ and $\ket{1,0}$. The ML analysis is designed to determine the populations (in the given basis) of the measurement qubit. Thus the references are designed to yield histograms associated with known populations in the measurement qubit. To first order, the reference states have all population in the measurement qubit. Contributions from the small populations outside the measurement qubit manifold are observationally equivalent to population in $\ket{\downarrow}$ since they are dark to a high degree.  If these populations are the same in all experiments, they are equivalent to a background contribution. With this in mind, we inspect the systematic effects from imperfect optical pumping and transfer in more detail.

Consider the effect of the $\epsilon\leq 1\times10^{-3}$ population (per ion) in states other than $\ket{2,2}$ after optical pumping (see next section). This population is distributed over the other states in a way that depends on details of the optical pumping process. With one exception considered below, population in states other than $\ket{2,2}$ is nominally dark in all reference histograms. The ML algorithm infers $q_j(c)$ as if all populations were in the measurement qubit manifold. Provided population in other state remains dark in all experiments, it is treated as a background and subtracted. The effect is that the algorithm infers the renormalized density matrix on the qubit rather than the actual one with trace reduced by the population outside the qubit manifold. This condition holds for our experiments: To first order, the $\epsilon$ population (of each ion) is outside of the computational qubit manifold during the Bell state preparation and outside the measurement qubit manifold. To correct for this effect and determine a lower bound on the overall Bell state fidelity, we subtracted $2\epsilon$ from the ML inferred fidelity. This is consistent with the effect on Bell state fidelity from simulations (see next section). The exception to this model is that a fraction of the $\epsilon$ non-$\ket{2,2}$ population after optical pumping is in the qubit manifold. In the case where non-$\ket{2,2}$ state is in the $\ket{\uparrow}$ state, reference histogram $r_3(c)$ and $r_4(c)$ are affected differently compared to the situation above. This is because in this case, the population stays inside the measurement qubit manifold. With respect to the background interpretation, this is equivalent to having the $\ket{2,2}$ population in references $r_3(c)$ and $r_4(c)$ exceed $0.5$ (for each ion) by $\xi\leq \epsilon/2$. To determine the effect on the ML inferred fidelity, we performed a sensitivity analysis on simulated data by varying the ML assumed populations for these references according to the parameter $\xi$. The change in fidelity is small compared to our uncertainties. Non-$\ket{2,2}$ population in the $\ket{\downarrow}$ state after optical pumping enters the qubit state at the beginning of Bell state preparation and in this context does not behave as a dark state independent of the analysis pulses. However, simulations of the optical pumping process show that this population is less than $10^{-4}$, which is small compared to our uncertainties. 

For the transfer of $\ket{2,2}$ to the $\ket{\uparrow}$ computational qubit state before the application of the gate, the errors of the transfer pulse result in population in the $\ket{2,2}$ state, outside the computational qubit manifold and unaffected by the MS interaction and the following analysis pulses. To first order, the transfer pulse after the gate will now move the population in the $\ket{2,2}$ state back to the $\ket{1,1}$ state, which will be nominally detected as dark, independent of the analysis pulses. While this is equivalent to a leakage error on the Bell state being analyzed, it is not accounted for by the ML analysis. Transfer error after the application of the gate results in extra dark population that depends on the final population in $\ket{1,1}$, which in turn depends on the analysis pulses. Thus, both effects are inconsistent with the analysis pulse model that is assumed by the ML analysis.  Such inconsistencies, if significant, are expected to show up in the bootstrapped likelihood-ratio test, but did not. Nevertheless, we performed a second sensitivity analysis on simulated data, where we modified the model to include an extra dark state outside the qubit manifold and included the expected transfer pulse effects by modifying the analysis pulses with pulses coupling the qubit to the extra state. The effect on the inferred error of this modification was also small compared to the uncertainties.

\subsection{Imperfect optical pumping and lower bound of Bell state fidelity}

The lower bound of the $\ket{2,2}$ state purity is deduced by deriving an upper bound on the error $\epsilon$ of preparing the $\ket{2,2}$ state after applying optical pumping. The population is dominantly in the $\ket{2,2}$ state but we do not know precisely which of the remaining hyperfine states are populated.  We write the density matrix of a single ion for this situation as

\begin{equation}
\rho = (1-\epsilon)\ket{2,2}\bra{2,2} + \sum_{i=1}^{7}\epsilon_i\ket{\Psi_i}\bra{\Psi_i},
\end{equation}
where $\epsilon = \sum_i^7\epsilon_i$ and $\ket{\Psi_i}$ represents the hyperfine states excluding the $\ket{2,2}$ state. One strategy for setting an upper bound on $\epsilon$ is to choose a cut-off count $\beta$ and compare the small-count ``tail'' probabilities $t=\sum_{c<\beta}h(c)$. 

Let $t_b$ be the tail probability of $h_{\ket{2,2}}$. Because state preparation and detection are not perfect, we have $t_b\geq\bar{t}_b$, where $\bar{t}_b$ is the tail probability of perfectly prepared $\ket{2,2}$ states. The tail probabilities $t_i$ for $\Psi_i$ are large, as verified by experimentally preparing each $\Psi_i$ state and measuring its count distribution.  From this we can set a lower bound on $t_i$ such that $t_i>l$. With this, we can write

\begin{eqnarray}
t_b&=&(1-\epsilon)\bar{t}_b + \sum_i^7\epsilon_i t_i\\
&\geq&(1-\epsilon)\bar{t}_b + \epsilon l\\
t_b&\geq&\epsilon\left(l-\bar{t}_b\right)+\bar{t}_b,
\end{eqnarray}
or
\begin{eqnarray}
\epsilon &\leq& \frac{t_b-\bar{t}_b}{l-\bar{t}_b}\\
&\leq& \frac{t_b}{l-\bar{t}_b}.
\end{eqnarray}

For our parameters of $l=0.8$ and $\bar{t}_b=0$ we estimate an upper bound on $\epsilon$ of $1\times10^{-3}$. We also numerically simulate the effect of imperfect optical pumping and find that the Bell state error scales linearly in as a function of $\epsilon$. This is consistent with the lower bound on overall Bell state fidelity of 0.997 inferred in the previous section by considering the effect on the ML analysis.

\subsection{Average Gate Fidelity}

To characterize the performance of the gate over all input states, we investigate the average gate fidelity, $F_{\mathrm{avg}}$ \cite{Horodecki1999,Nielsen2002}, by employing numerical simulation with known experimental imperfections. Firstly, we write

\begin{eqnarray}
F_{\mathrm{avg}}=\frac{6}{5}S_{+} +\frac{3}{5}S_{-} - \frac{1}{5},
\label{AverageFidelity}
\end{eqnarray}
with
\begin{widetext}
\begin{eqnarray}
S_+ &=& \frac{1}{36}\sum_{U_1}\sum_{U_2}\left[\left(\bra{U_1}\otimes\bra{U_2}\right)\hat{G}_{\mathrm{ideal}}^{\dagger}\rho_{\mathrm{noisy}}(U_1,U_2)\hat{G}_{\mathrm{ideal}}\left(\ket{U_1}\otimes\ket{U_2}\right)\right],\label{EqSplus}\\
S_- &=& \frac{1}{36}\sum_{U_1}\sum_{U_2}\left[\left(\bra{\overline{U_1}}\otimes\bra{\overline{U_2}}\right)\hat{G}_{\mathrm{ideal}}^{\dagger}\rho_{\mathrm{noisy}}(U_1,U_2)\hat{G}_{\mathrm{ideal}}\left(\ket{\overline{U_1}}\otimes\ket{\overline{U_2}}\right)\right].\label{EqSMinus}
\end{eqnarray}
\end{widetext}
where $\ket{U_i}$ is an eigenstates of Pauli operators $\hat{\sigma}_x$, $\hat{\sigma}_y$ or $\hat{\sigma}_z$ for the $i$th qubit, and $\ket{\overline{U_i}}$ is the state orthogonal to $\ket{U_i}$. We fix a consistent phase for these eigenstates throughout. The operator $\hat{G}_{\mathrm{ideal}}$ is the ideal entangling operation, $\rho_{\mathrm{noisy}}(U_1,U_2)$ represents the resultant density matrix of the imperfect entangling operation with the input states of $\ket{U_1}\otimes\ket{U_2}$.

Equation (\ref{AverageFidelity}) can be verified by direct computation, or by noting that it is invariant under one-qubit Clifford operations and SWAP. There are three independent such invariant expressions, so it suffices to check validity on a small number of simple quantum operations. Definitions and expression for $F_{\mathrm{avg}}$ can be found in \cite{Horodecki1999,Nielsen2002}. We use Eq. (\ref{AverageFidelity}) to compute $F_{\mathrm{avg}}$ instead of alternative expressions \cite{Horodecki1999,Pedersen2007} in order to bypass computation of an explicit process matrix. With 36 different input states, our simulations of known imperfections yield the summands in Eq. (\ref{EqSplus}) and Eq. (\ref{EqSMinus}). We found $F_{\mathrm{avg}}$ lies within the uncertainty of the inferred Bell state fidelity measurement.

\subsection{Error sources}

Spontaneous emission error is caused by randomly scattered photons when driving stimulated Raman transitions. It can be separated into Raman and Rayleigh scattering. Raman scattering processes are inelastic and project an ion's internal state to one of the other hyperfine states, destroying coherence. Rayleigh scattering processes are elastic and do not necessarily cause spin decoherence \cite{Ozeri2007}; however, momentum kicks from photon recoil cause uncontrolled displacements of the motional state, which result in phase errors in the final states. Raman scattering can be reduced by increasing $|\Delta|$ at the cost of higher laser intensity to maintain the same Rabi rate. However, Rayleigh scattering error cannot be reduced by increasing the detuning and it reaches an asymtoptic value. This error is proportional to the Lamb-Dicke parameter and thus could be reduced by increasing the trap frequency; it can also be reduced by using multiple loops in phase space \cite{Ozeri2007,Hayes2012}. These methods reduce the gate Rabi rate and thus increase Raman scattering error. In our experiment, eliminating the axial micromotion would allow us to increase $\Delta$ by a factor of $\xi \simeq$ 2 which would lower the Raman scattering error by a factor of $2\xi$, and the Rayleigh scattering error by a factor of $\xi$ while maintaining the same gate duration.

Spontaneous Raman scattering can result in leakage of population from the qubit manifold. The resulting states will predominantly be detected as dark and falsely associated with the qubit $\ket{\downarrow}$ state. This creates a systematic bias that overestimates the actual Bell state fidelity. Through simulations, we found that such a bias is approximately $4\times10^{-5}$ for the Bell state fidelity created at a Raman detuning of $-2\pi\times 900$ GHz and approximately $1.5\times 10^{-3}$ for $-2\pi\times 90$ GHz Raman detuning. 

Motional mode frequency fluctuations also cause errors. For the stretch mode, the sources of frequency fluctuations (which are slow compared to the gate durations shown in Fig. \ref{fig:GateDuration}) are (i) fluctuations in the DC potentials applied to electrodes for trapping, (ii) fluctuating electric-field gradients from uncontrolled charging of electrode surfaces \cite{Harlander2010}, and (iii) non-linear coupling to transverse ``rocking'' modes \cite{Roos2008,Nie2009}. By measuring the lineshape for exciting the motional state of a single ion with injected RF ``tickle" potentials on the trap electrodes at frequencies near the mode frequencies, we estimate the first two sources contribute fluctuations of approximately 50 Hz. Stray charging can be caused by UV beam light scattering off the trap surfaces so this effect may becomes more pronounced when higher laser power is used. For (iii), to a good approximation, the shift of the stretch mode frequency from excitation of the rocking modes is given by $\delta \omega_S = \chi(n_x + n_y +1)$  where $\chi$ is a non-linear coupling parameter and $n_x$ and $n_y$ are Fock state occupation numbers of the two transverse rocking modes \cite{Roos2008,Nie2009}. For our parameters $\chi \simeq 45$ Hz, our Raman laser beam geometry did not allow direct measurement of the $\langle n_x\rangle$ and $\langle n_y\rangle$ radial modes excitation. Therefore, the final temperature is estimated from the (thermal) Doppler cooling limit, taking into account heating due to photon recoil during sideband cooling of the axial modes. From this, we estimate the stretch mode frequency fluctuations from experiment to experiment to be approximately 100 Hz r.m.s. As these fluctuations are dependent on the occupation numbers of the radial modes, the error can be suppressed by cooling the radial modes to the ground state.  In Fig. \ref{fig:GateDuration}, we show three simulation curves for different total values of the r.m.s. frequency fluctuation of the motional mode that follow the trend of the data and are consistent with our known sources. The error due to these fluctuations is approximately $1 \times 10^{-4}$ for the shortest gate durations.

Errors are also caused by changes in the MS Rabi rates, which cause fluctuations in the state-dependent forces.  Sources are (i) fluctuations of the ions' micro-motion amplitude along the axial direction, (ii) fluctuations in the laser beam intensities at the ion locations, and (iii) fluctuations in the Debye-Waller factor associated with the center-of-mass (C) mode \cite{bible}. In our experiments, the latter gives the largest error.  Through numerical simulation, we derived the expression $\epsilon \simeq 2.5\times (\frac{\delta\Omega}{\Omega})^2$ describing the MS gate error due to Rabi rate fluctuations (this agrees with the expression in \cite{Benhelm2008}).  Given a thermal distribution of the C mode, the r.m.s. Rabi rate fluctuation of the stretch mode can be described by $\langle\frac{\delta\Omega}{\Omega}\rangle = \eta_{\mathrm{C}}^2\sqrt{\langle n_{\mathrm{C}} \rangle(\langle n_{\mathrm{C}}\rangle +1)}$ where $\eta_{\mathrm{C}}$ is the Lamb-Dicke parameter for that mode \cite{bible}. With  $\langle n_{\mathrm{C}}\rangle \simeq$ 0.01 at the beginning of the MS interaction, we find $\langle\frac{\delta\Omega}{\Omega}\rangle \simeq 6 \times 10^{-3}$ and we deduce an error of approximately $1\times 10^{-4}$ to $\ket{\Phi_+}$. Because this mode will experience anomalous heating during the gate, the actual error contribution will increase with the gate duration.  The heating rate for the COM mode is approximately 80 quanta per second. For our 30 $\mu$s gate duration, this implies a change of $\Delta \langle n_C \rangle \simeq 0.001$ averaged over the duration of the gate.  Therefore the error caused by the modification of the Debye-Waller factor from heating can be neglected for our fastest gate times.  

Because the two-qubit-gate Raman transitions are driven on the second micro-motion sideband, the Rabi rates are proportional to the second-order Bessel funtion $J_2(|\bm{\Delta k}| z_{\mu m}) \simeq 0.48$, where $|\bm{\Delta k}| z_{\mu m} = 2.9$ is the modulation index due to the micromotion-induced Doppler shift and is proportional to the applied RF voltage $V_{\mathrm{RF}}$. For the conditions of the experiment, $J_2(|\bm{\Delta k}| z_{\mu m})$ is near a maximum such that the Rabi rate is relatively insensitive to fluctuations in $V_{\mathrm{RF}}$. Our measurements show that the transverse mode frequencies can drift by up to 10 kHz over the course of several experiments; this would imply a relative drift in $V_{\mathrm{RF}}$ of $\sim 1 \times 10^{-3}$ and a corresponding change in the Rabi rate of $3 \times 10^{-4}$, which contributes an error that is negligible compared to the other errors. 

Laser intensity fluctuations can be assumed to be comparable to the fluctuations measured from the single-qubit benchmarking experiments ($ \sim 1 \times 10^{-3}$), which makes this contribution to Rabi rate fluctuations negligible compared to that of the fluctuating Debye-Waller factors. Laser intensity fluctuations also cause fluctuation in AC-Stark shifts, which we measure to be $\sim$ 1 kHz at a Raman detuning of $-2\pi\times 900$ GHz and induce negligible error. 

Smaller sources of error are (i) laser beam phase fluctuations between beam paths 1 and 2 during each experiment, (ii) individual qubit decoherence, (iii) heating of the axial stretch mode \cite{Turchette2000}, (iv) imperfect Lamb-Dicke approximation, and (v) off-resonance coupling to spectator transitions. Each of these sources contributes a few times $10^{-5}$ error to the entangling gate.

Sources of frequency and phase fluctuations include fluctuations in the laser beam phases $\phi_{j,b}$ and $\phi_{j,r}$, and fluctuations in the qubit frequency. Fluctuations due to relative length changes between paths 1 and 2 were measured by recombining the two beams after they exit the UV fibers, detecting with a fast photo-diode,  and measuring the phase of the beat note using the AOM RF sources as a reference.  We measured a phase drift of $\sim \pi$ after $\sim$ 1 s, this is likely due to temperature drift of the optical elements in the setup. We also observed small-amplitude phase oscillations with frequencies of a few hundred Hertz, which can be attributed to acoustic vibrations in the laboratory. With this, we estimate an error of $\sim 2\times10^{-5}$ to the gate. The measured coherence time of the qubit from Ramsey experiments is approximately 1.5 s, which implies an r.m.s. qubit transition frequency error of 1 Hz, giving negligible error compared to other sources. 

The heating rate of the axial stretch mode is measured to be less than 1 quantum per second and contributes an error of less than $3\times10^{-5}$ to $\ket{\Phi_+}$. The M$\o$lmer-S$\o$rensen interaction is robust against finite thermal excitation in the Lamb-Dicke limit, $\eta \ll 1$. However, due to the small mass of $^9$Be$^+$ ions this condition is not rigorously satisfied and the sensitivity to finite motional excitation must be considered. The error due to this is given by $\frac{\pi^2}{4}\eta^4\langle n \rangle(\langle n \rangle+1)$ \cite{Sorensen2000}, which corresponds to an error of less than $2\times10^{-5}$ for our parameters. We also use numerical simulation to study this effect and find good agreement. Even within the Lamb-Dicke limit, finite thermal excitation increases the sensitivity of error due to motional mode frequency fluctuations \cite{Hayes2012}. For our parameters, this error is negligible.

Off-resonant coupling to spectator transitions is suppressed by employing laser pulse shaping. The rise and fall durations of the gate pulse are adjusted such that the Fourier component at the frequencies of spectator transitions is sufficiently small. Spectator transitions include the carrier and COM sideband transitions as well as other atomic transitions that can be coupled by micromotion sidebands (the Zeeman splittings between atomic states are comparable to $\omega_{\mathrm{RF}}$). If a square pulse is used instead of a shaped pulse, we estimate an error of $1\times10^{-4}$ for a gate duration of 30 $\mu $s \cite{Sorensen2000}.


\begin{thebibliography}{10}

\bibitem{Feynman1982}
R. P. Feynman, Int. J. Theor. Phys. {\bf 21}, 467  (1982).

\bibitem{Deutsch1985}
D. Deutsch, Proc. R. Soc. Lond. A {\bf 400}, 97 (1985).

\bibitem{Lloyd1996}
S. Lloyd, Science {\bf 273}, 1073-1078 (1996).

\bibitem{Preskill1998}
J. Preskill, Proc. R. Soc. Lond. A {\bf 454},  385  (1998).

\bibitem{Knill2010}
E. Knill, Nature {\bf 463},  441  (2010).

\bibitem{Ladd2010}
T. D. Ladd, F. Jelezko, R. Laflamme, Y. Nakamura, C. Monroe, and J. L. O'Brien, Nature {\bf 464},  45  (2010).

\bibitem{Home2009}
J. P. Home, D. Hanneke, J. D. Jost, J. M. Amini, D. Leibfried, and D. J. Wineland, Science {\bf 325}, 1227 (2009).

\bibitem{Hanneke2010}
D. Hanneke, J. P. Home, J. D. Jost, J. M. Amini, D. Leibfried, and D. J. Wineland, Nature Physics {\bf 6}, 13 -16 (2010).

\bibitem{Bowler2012}
R. Bowler, J. Gaebler, Y. Lin, T. R. Tan, D. Hanneke, J. D. Jost, J. P. Home, D. Leibfried, and D. J. Wineland, Phys. Rev. Lett. {\bf 109}, 080502 (2012).

\bibitem{Walther2012}
A. Walther, F. Ziesel, T. Ruster, S. T. Dawkins, K. Ott, M. Hettrich, K. Singer, F. Schmidt-Kaler, and U. Poschinger, Phys. Rev. Lett. {\bf 109}, 080501 (2012).

\bibitem{Ozeri2007}
R. Ozeri, W. M. Itano, R. B. Blakestad, J. Britton, J. Chiaverini, J. D. Jost, C. Langer, D. Leibfried, R. Reichle, S. Seidelin, J. H. Wesenberg, and D. J. Wineland, Phys. Rev. A {\bf 75},  042329  (2007).

\bibitem{Benhelm2008}
J. Benhelm, G. Kirchmair, C. F. Roos, and R. Blatt, Nat. Phys {\bf 4}, 463 (2008).

\bibitem{Ballance2015}
C. J. Ballance, T. P. Harty, N. M. Linke, M. A. Sepiol and D. M. Lucas, arXiv:1512.04600.

\bibitem{Blakestad2011}
R. B. Blakestad, C. Ospelkaus, A. P. VanDevender, J. H. Wesenberg, M. J. Biercuk, D. Leibfried, and D. J. Wineland, Phys. Rev. A {\bf 84}, 032314 (2011).

\bibitem{bible}
D. J. Wineland, C. Monroe, W. M. Itano, D. Leibfried, B. E. King, and D. M. Meekhof, J. Res. Natl. Inst. Stand. Technol. {\bf 103},  259  (1998).

\bibitem{Kielpinski2002}
D. Kielpinski, C. R. Monroe, and D. J. Wineland, Nature {\bf 417}, 709 - 711 (2002).

\bibitem{Langer2005}
C. Langer, R. Ozeri, J. D. Jost, J. Chiaverini, B. DeMarco, A. Ben-Kish, R. B. Blakestad, J. Britton, D. B. Hume, W. M. Itano, D. Leibfried, R. Reichle, T. Rosenband, T. Schaetz, P. O. Schmidt, and D. J. Wineland, Phys. Rev. Lett. {\bf 95},  060502  (2005).

\bibitem{Monroe1995}
C. Monroe, D. M. Meekhof, B. E. King, S. R. Jefferts, W. M. Itano, D. J. Wineland, and P. Gould, Phys. Rev. Lett. {\bf 75}, 4011 (1995).

\bibitem{Colombe2014}
Y. Colombe, D. H. Slichter, A. C. Wilson, D. Leibfried, and D. J. Wineland, Opt. Express {\bf 22}, 19783 (2014).

\bibitem{Brown2011}
K. R. Brown, A. C. Wilson, Y. Colombe, C. Ospelkaus, A. M. Meier, E. Knill, D. Leibfried, and D. J. Wineland, Phys. Rev. A {\bf 84}, 030303(R) (2011).

\bibitem{Sorensen1999}
A. S\o{}rensen and K. M\o{}lmer, Phys. Rev. Lett. {\bf 82},  1971  (1999).

\bibitem{Sorensen2000}
A. S\o{}rensen and K. M\o{}lmer, Phys. Rev. A {\bf 62},  022311  (2000).

\bibitem{Milburn1999}
G. J. Milburn, S. Schneider, and D. F. V. James, Fortschr, Phys. {\bf 48}, 801 (2000).

\bibitem{Solano1999}
E. Solano, R. L. de Matos Filho, and N. Zagury, Phys. Rev. A {\bf 59},  R2539  (1999).

\bibitem{Efron1993} 
B. Efron, and R. J. Tibshirani, An introduction to the bootstrap (Chapman and Hall, 1993).

\bibitem{Turchette2000}
Q. A. Turchette, D. Kielpinski, B. E. King, D. Leibfried, D. M. Meekhof, C. J. Myatt, M. A. Rowe, C. A. Sackett, C. S. Wood, W. M. Itano, C. Monroe, and D. J. Wineland, Phys. Rev. A {\bf 61}, 063418 (2000).

\bibitem{Gaebler2012}
J. P. Gaebler, A. M. Meier, T. R. Tan, R. Bowler, Y. Lin, D. Hanneke, J. D. Jost, J. P. Home, E. Knill, D. Leibfried, and D. J. Wineland, Phys. Rev. Lett. {\bf 108},  260503  (2012).

\bibitem{Navon2014}
N. Navon, N. Akerman, S. Kotler, Y. Glickman, and R. Ozeri, Phys. Rev. A. {\bf 90}, 010103(R) (2014).

\bibitem{Wilson2011}
A. C. Wilson, C. Ospelkaus, A. Vandervender, J. A. Mlynek, K. R. Brown, D. Leibfried, and D. J. Wineland, Appl. Phys. B {\bf 105}, 741 (2011).

\bibitem{Leibrandt2015}
D. R. Leibrandt and J. Heidecker, Rev. Sci. Instrum. {\bf 86}, 123115 (2015).

\bibitem{Bowler2013}
R. Bowler, U. Warring, J. W. Britton, and J. M. Amini, Rev. Sci. Instrum. {\bf 84}, 033108 (2013).

\bibitem{Levitt1986}
M. H. Levitt, Prog. NMR Spectrosc. {\bf 18}, 61 (1986).

\bibitem{Lee2005}
P. J. Lee, K.-A. Brickman, L Deslauriers, P. C. Haljan, L.-M. Duan, and C. Monroe, J. Opt. B {\bf 7}, S371 (2005).

\bibitem{Tan2015}
T. R. Tan, J. P Gaebler, Y. Lin, Y. Wan, R. Bowler, D. Leibfried, and D. J. Wineland, Nature {\bf 528}, 380 (2015).

\bibitem{Hradil2004}
Z. Hradil, J. \v{R}eh\'{a}\v{c}ek, J. Fiur\'{a}\v{s}ek, M. Je\v{z}ek, {\it Quantum state estimation} (Springer, New York 2004), pp. 59-100. 

\bibitem{Boos2003}
D. D. Boos, Statistical Science {\bf 18}, 168 (2003).

\bibitem{Lin2016}
Y. Lin, J. P. Gaebler, F. Reiter, T. R. Tan, R. Bowler, Y. Wan, A. Keith, E. Knill, S. Glancy, K. Coakley, A. S. S\o{}rensen, D. Leibfried, and D. J. Wineland, arXiv:1603.03848.

\bibitem{Horodecki1999}
M. Horodecki, P. Horodecki, and R. Horodecki, Phys. Rev. A {\bf 60}, 1888 (1999).

\bibitem{Nielsen2002}
M. A. Nielsen, Phys. Lett. {\bf 303}, 249-252 (2002).

\bibitem{Pedersen2007}
L. H. Pedersen, N. M. M\o{}ller and K. M\o{}lmer, Phys. Lett. A {\bf 367}, 47-51 (2007).

\bibitem{Hayes2012}
D. Hayes, S. M. Clark, S. Debnath, D. Hucul, I. V. Inlek, K. W. Lee, Q. Quraishi, and C. Monroe, Phys. Rev. Lett. {\bf 109}, 020503 (2012). 

\bibitem{Harlander2010}
M. Harlander, M. Brownutt, W. Hansel, and R. Blatt, New J. Phys. {\bf 12}, 093035 (2010).

\bibitem{Roos2008}
C. F. Roos, T. Monz, K. Kim, M. Riebe, H. H\"{a}ffner, D. F. V. James, and R. Blatt, Phys. Rev. A {\bf 77}, 040302(R) (2008).

\bibitem{Nie2009}
X. R. Nie, C. F. Roos, and D. F. V. James, Phys. Letts. A {\bf 373}, 422-425 (2009).

\end{thebibliography}
\end{document}